\documentstyle[preprint,prl,aps]{revtex}
\begin{document}
\draft
\preprint{RU-94-94}
\title{Fermi and Non-Fermi Liquid behavior in the anisotropic
Multichannel Kondo Model -- Bethe Ansatz solution}
\author{Natan Andrei and  Andr\'es Jerez}
\address{Department of Physics and Astronomy, Rutgers University,
Piscataway, NJ 08855}
\maketitle

\begin{abstract}
We solve the Multichannel Kondo model with channel anisotropy using
the Bethe Ansatz method. The model generates  energy
scales, characterizing the neighborhoods of the various infrared
fixed
points, reflecting the structure of the symmetry breaking
in the channel sector.
The nature of these fixed points also depends on the magnitude of
the impurity spin $S$. We present a detailed discussion for the two
channel case and point out some new non-Fermi liquid behavior.
\end{abstract}

\pacs{75.20.Hr}

\narrowtext

Non-Fermi liquid (NFL) behavior observed  in
some Ce and U alloys \cite{maple} has stimulated an intense study
of
many impurity
models; in particular, of the multichannel
Kondo model \cite{noziers80}:
\begin{eqnarray}
H =  H_0 + 2 \sum_{a,b} \sum_{m=1}^{f}
J_m\psi_{a,m}^{\dagger}(0)
\mbox{\boldmath $\sigma$}_{ab}\psi_{b,m}(0)\cdot {\bf S},
\label{1}
\end{eqnarray}
where $H_0= -i \sum_{a; m} \int dx
\psi_{a,m}^{\dagger}(x) \partial_x \psi_{a,m}(x)$.
Here the field $\psi_{a,m}$ describes electrons with spin index $a
=
\pm 1/2$ and  orbital channel ({\it flavor}) index $m,\;
m=1...f$, and  we chose to set $v_F=1$. The operator $\bf S$ represents the
spin-$S$ impurity
localized at $x=0$.

The infrared behavior
 of the model
depends on the structure of the flavor sector \cite{noziers80}.
 In particular,
a new  behavior appears when $f> 2S$, with the overscreened
system exhibiting NFL physics.

An exact solution was found for the  isotropic
case, $J_m=J$ \cite{adkondo,wt}. The solution yields the spectrum and
the thermodynamics for any temperature and magnetic field.
In particular, the residual entropy and the critical exponents
 governing the low-energy physics were calculated.
The thermodynamics was studied numerically in
\cite{schlot}.

The neighborhood of the isotropic fixed point was
further studied by means of Conformal Field Theory \cite{al} and
Bosonization methods
\cite{ek}, yielding the long distance asymptotics of the
correlation
functions with  critical exponents
that are same as those characterizing the thermodynamic functions.

In this paper we
present the solution of the  channel-anisotropic model.
We shall discuss in detail the case of two channels and arbitrary
spin $S$, and briefly
outline the generalization for more flavors.

We shall find that in the two channel case the model
generates two scales,
$T_i$ and $T_a$, which we shall later interpret as associated with
two fixed points, the isotropic and anisotropic,
respectively. The
scales are explicitly given (in our cutoff scheme) by
$
T_i  \equiv D e^{-\frac{\pi}{J_1}},~~
T_a  \equiv D \cos \left( \frac{J_1}{J_2}\frac{\pi}{2} \right)
e^{-\frac{\pi}{J_2} },
$
where $D=N/L$. Here $N$ is the number of electrons in each channel,
$L$ is the
length of the system, and $J_1\leq J_2$. In the scaling limit
 $D$ is taken to infinity with
the (bare) couplings' dependence on $D$ chosen so as
to keep the scales finite. The functional
dependence of the scales on the  coupling constants $J_1,
J_2$, is not universal
and may change with the cutoff procedure. However, the
dependence of physical quantities  on the scales is universal.
The ratio $\Delta \equiv T_a/T_i$
is the physical measure of the anisotropy.


The presence of flavor allows up to $f$ electrons to interact
simultaneously with the impurity. Therefore, spin
composites of electrons form
irrespectively of the degree of anisotropy, and their  binding
energy  is set by the smallest of the couplings. The
hamiltonian must be regularized with care to allow the formation of the
composites while maintaining integrability. We choose the regularization
scheme used in \cite{adkondo}, leading to
the (regularized) first quantized form of the hamiltonian
(\ref{1}),
\begin{eqnarray}
h = \sum_{j=1}^{2N} \left( -i\partial_j - \Lambda^{-1}
(\partial_j)^{2} + 2J_j\delta(x_j) \mbox{\boldmath  $\sigma$}_j\cdot
{\bf S}\right).
\label{2}
\end{eqnarray}
Here $J_j$ is either $J_1$ or $J_2$. The limit
$\Lambda \rightarrow \infty$ will be taken  only after
determining the eigenvalues. This limiting procedure, the
 {\it fusion} \cite{adkondo}, leads to the formation
of the spin composites. We shall find that the
binding energy will be well above the spin
energy scales, and the flavor excitations will disappear
from the low-energy  spectrum.

The eigenfunctions of (\ref{2}) are combinations of plane waves with
pseudomomenta $\{ k_j,
j=1,\dots,2N\}$ and amplitudes
$A_{a_1....a_{2N};\alpha
}^{m_{1}...m_{2N}}$
depending on the electron spin and flavor indices {$a_j,m_j$},
and the impurity spin index $\alpha$.
The energy eigenvalues in terms of the pseudomomenta are
$
E = \sum_{j=1}^{2N} k_j(1+k_j/\Lambda)
$,
while the amplitudes are determined from
the two body S-matrices,
to
which we now turn.

The impurity-electron  S-matrix is  derived from (\ref{2}),
\begin{eqnarray}
S_{j0} &=& \frac{\tilde{\lambda}_j + 1 -
iJ_j\left(\mbox{\boldmath $\sigma$}\cdot{\bf S} +
\frac{1}{2}\right)} {\tilde{\lambda_j} +1 -iJ_j}  ,
\label{4}
\end{eqnarray}
where $\tilde{\lambda}_j  = k_j/ \Lambda$.
Since the interactions are flavor preserving, the S-matrix
(\ref{4})
has only a non-trivial term in spin space.

All electrons are
right-movers
and as there is no direct interaction between them the
electronic states are infinitely degenerate away from the impurity.
This permits the
introduction of an arbitrary electron-electron S-matrix $S_{jl}$ into the
definition of the
wave-functions. To construct a
basis that will manifest the integrability of the model
 the matrices $S_{jl}$ must be chosen so as to satisfy the
Yang-Baxter
factorization equations,
\begin{eqnarray}
S_{j0}S_{l0}S_{jl}  =  S_{jl}S_{l0}S_{j0},~~~
S_{jk}S_{lk}S_{jl}  =  S_{jl}S_{lk}S_{jk}.
\label{5}
\end{eqnarray}
The solution for $S_{jl}$ can be given as a direct
product of spin and flavor
terms, $S_{jl} = S_{jl}^{spin} \otimes S_{jl}^{flavor}$,
with each term satisfying (\ref{5}) separately.
The
spin component of $S_{jl}$ is given by
\begin{eqnarray}
S_{jl}^{spin} = \frac{\left( \lambda_j+\frac{1}{J_j}\right) -
\left( \lambda_l+\frac{1}{J_l}\right) - iP_{jl}}
{\left( \lambda_j+\frac{1}{J_j}\right) -
\left( \lambda_l+\frac{1}{J_l}\right) - i},
\label{7}
\end{eqnarray}
where $\lambda_j=\tilde{\lambda}_j / J_j$ and $P_{jl} = \frac{1}{2}
\left(\mbox{\boldmath $\sigma$}_j\cdot\mbox{\boldmath
$\sigma$}_l+1\right)$. The form of
(\ref{7}) reflects the $SU(2)_{spin}$ invariance
of the model.
The  flavor component of $S_{jl}$ reflects the breaking of the
$SU(2)_{flavor}$ symmetry to a residual $U(1)$ when the anisotropy
is present, and is given by \cite{fad}
\begin{eqnarray*}
\lefteqn{S_{jl}^{flavor} = \frac{i}{2} \frac{\sin
\nu}{\sinh(\kappa(\tilde{\lambda_j}-\tilde{\lambda_l})+i\nu)}
\left\{ \tau_x \otimes \tau_x + \tau_y \otimes \tau_y \right\}}
\nonumber \\
&&+\frac{1}{2}\{1 + \tau_z \otimes \tau_z\}
+\frac{1}{2}\frac{\sinh
(\kappa(\tilde{\lambda}_j - \tilde{\lambda}_l))}{\sinh
(\kappa(\tilde{\lambda}_j - \tilde{\lambda}_l)+i\nu)}\{1-
\tau_z \otimes \tau_z\}.
\end{eqnarray*}
Here, $\{\tau \}$ are the Pauli matrices, and $\kappa$ and $\nu$
are functions of  the couplings. Denoting $\mu\equiv \nu/\kappa$,
we shall see that $\mu$ is related to the binding energy of the
composites.

Given the set of consistent S-matrices we may derive the
Bethe Ansatz equations determining the allowed pseudomomenta $k_j$
and hence the spectrum. Introducing the  auxiliary variables
$\{\omega_\gamma\}$ for the flavor sector, and $\{\chi_\gamma\}$
for the spin sector, we find:
\begin{eqnarray*}
&&e^{ik_jL} = \prod_{\delta=1}^{M}
\frac{\chi_\delta-\lambda_j-\frac{1}{J_j} + \frac{i}{2}}
{\chi_\delta-\lambda_j-\frac{1}{J_j} - \frac{i}{2}}
\prod_{\delta=1}^{M_1}
\frac{\sinh(\nu(\omega_\delta-\frac{\tilde{\lambda}_j}{\mu}+\frac
{i}{2}))}
{\sinh(\nu(\omega_\delta-\frac{\tilde{\lambda}_j}{\mu}-\frac{i}{2
}))}, \\
&&\prod_{j=1}^{2N}
\frac{\sinh(\nu(\omega_\gamma-\frac{\tilde{\lambda}_j}{\mu}+\frac
{i}{2}))}
{\sinh(\nu(\omega_\omega-\frac{\tilde{\lambda}_j}{\mu}-\frac{i}{2
}))}
= -\prod_{\delta=1}^{M_1}
\frac{\sinh(\nu(\omega_\gamma-\omega_\delta+\frac{i}{2}))}
{\sinh(\nu(\omega_\gamma-\omega_\delta-\frac{i}{2}))}, \\
&&-\prod_{\delta=1}^{M} \frac{\chi_\gamma-\chi_\delta+i}
{\chi_\gamma-\chi_\delta-i} = \frac{\chi_\gamma+iS}{\chi_\gamma-iS}
\prod_{j=1}^{2N}
\frac{\chi_\gamma-\lambda_j-\frac{1}{J_j}+\frac{i}{2}}
{\chi_\gamma-\lambda_j-\frac{1}{J_j}-\frac{i}{2}},
\end{eqnarray*}
describing the full content of the model.

The ground state and low lying energy excitations reside in the
flavor-singlet sector where the solutions are in the form of {\it
double 2-strings},
\begin{eqnarray}
\lambda_\delta^{++} &= &\frac{\mu}{J_2} (
\omega_\delta+\frac{i}{2}), \;
\lambda_\delta^{-+} = \frac{\mu}{J_1} (
\omega_\delta-\frac{i}{2}), \nonumber \\
\lambda_\delta^{+-} &=& \frac{\mu}{J_2} (
\omega_\delta-\frac{i}{2}),\;
\lambda_\delta^{--} = \frac{\mu}{J_1} (
\omega_\delta+\frac{i}{2}).
\label{14}
\end{eqnarray}
The structure of the double string  solution  reflects the flavor
symmetry breaking. In the isotropic limit the two strings coalesce
leading to the $SU(2)$ flavor degeneracy and the NFL-behavior.

The energy associated with the double 2-string is
$\varepsilon_\delta =
2\mu\omega_\delta \left(1+\mu\omega_\delta/\Lambda\right) -
\mu^2\Lambda/2$,
leading to  the
{\it 1-string} hypothesis for the $\{\omega_\gamma\}$, namely $
\omega_\delta = p_\delta/(\mu \Lambda)$ \cite{adkondo}.
We substitute the {\it double} 2-string solution
(\ref{14}) in the eigenvalue equations and  take the
$\Lambda\rightarrow
\infty$ limit. As explained, this
 is the {\it fusion} process in which the flavor
degrees of
freedom are removed from the low energy effective spin problem. We
find that the spin contribution to the energy is $
E = \sum_{\delta=1}^{\frac{N}{4}} 4 p_\delta$,
and the spin degrees of freedom are described by the following
system of fused Bethe Ansatz equations:
\begin{eqnarray*}
\lefteqn{-\prod_{\delta=1}^{M} \frac{\chi_\gamma-\chi_\delta+i}
{\chi_\gamma-\chi_\delta-i} =
\frac{\chi_\gamma+iS}{\chi_\gamma-iS}} \nonumber \\
&&\times \left[  \left(
\frac{\chi_\gamma-\frac{1}{J_2}+\frac{i}{2}(1+\phi)}
{\chi_\gamma-\frac{1}{J_2}-\frac{i}{2}(1+\phi)} \right)\left(
\frac{\chi_\gamma-\frac{1}{J_2}+\frac{i}{2}(1-\phi)}
{\chi_\gamma-\frac{1}{J_2}-\frac{i}{2}(1-\phi)} \right)  \right.
\nonumber \\  && \times \left.
\left( \frac{\chi_\gamma-\frac{1}{J_1}+\frac{i}{2}(1+\varphi)}
{\chi_\gamma-\frac{1}{J_1}-\frac{i}{2}(1+\varphi)} \right)\left(
\frac{\chi_\gamma-\frac{1}{J_1}+\frac{i}{2}(1-\varphi)}
{\chi_\gamma-\frac{1}{J_1}-\frac{i}{2}(1-\varphi)} \right)
\right]^{\frac{N}{2}},
\end{eqnarray*}
with $\phi \equiv  \mu/J_2$ and $\varphi \equiv
\mu/J_1$. As the binding energy is set by the weakest coupling, $\mu$
is related to $J_1$, and in
 the scaling limit we find $\mu =
J_1,~\phi=J_1/J_2,~\varphi=1$.  Note that the equations reduce to
the isotropic equations for $\phi=J_1/J_2=1$ and to
the 1-channel Kondo equations for $\phi=0,~J_1=0$.

We  now solve the equations, identify the ground state and
excitations, and summing over the latter, derive the free energy.
The main results are:

$(i)$ The solutions of the equations are of the form given by
the {\it string hypothesis} valid in the thermodynamic limit, $
\chi_\delta^{n,k} = \chi_\delta^n + \frac{i}{2}(n+1 -
2k),~~~k=1,\dots,n,~~\chi_\delta^n~\mbox{real}$.

$(ii)$ The ground state is  composed of $\chi$ 1- and 2-strings,
interpolating between the isotropic model (where $\phi=1$ and the
ground state built of 2-strings leading to NFL physics), and the
single channel model (where
$\phi=0$ and the ground state is built of 1-strings describing a
Fermi liquid (FL)).

$(iii)$  The impurity free energy for impurity spin-$S$,
anisotropy $\Delta$,  temperature $T$ and magnetic field $h$
is
\begin{eqnarray*}
F^i(\Delta, S; T,h) = -\frac{T}{2\pi} \int_{-\infty}^{\infty} d\xi
\frac{\ln
(1+\eta_{2S}(\xi,\frac{h}{T}))}{\cosh(\xi+\ln \frac{T}{T_i})}.
\end{eqnarray*}
The function $\eta_{2S}(\xi,h/T)$
belongs to the set of functions \{$\eta_n$\}
satisfying the following set of coupled integral equations, here
written in the scaling limit ($D \rightarrow \infty$, keeping $T_i,
T_a$
fixed),
\begin{eqnarray*}
\ln \eta_1 &=&- 2 \Delta e^\xi +
G\ln(1+\eta_2) \\
\ln \eta_2
&=&- e^{\xi}
+ G\ln(1+\eta_1)+G\ln(1+\eta_3)   \\
\ln \eta_n &=& G\ln(1+\eta_{n-1})+G\ln(1+\eta_{n+1}),~~~n > 2,
\end{eqnarray*}
with boundary conditions
$
\lim_{n\rightarrow \infty}
([n+1]\ln(1+\eta_n)-[n]\ln(1+\eta_{n+1})) =
-2\mu h/T.
$
The integral operators $[\alpha]$ and $G$ are defined by the
kernels $\alpha/(2\pi((\xi'-\xi)^2+(\frac{\alpha}{2})^2))$
and $1/(2\cosh (\pi
(\xi - \xi ')))$, respectively.

We now elaborate on $(iii)$. We studied analytically the equations in the
limits of small and large
anisotropy, both for $S=1/2$ and for $S=1$, and
deduced the infrared behavior of the thermodynamic functions and
its dependence on the anisotropy $\Delta$:

The $S=1/2$ impurity contribution to
the free energy at low temperatures is given by
\begin{eqnarray*}
F^i_{S=\frac{1}{2}} \sim \left\{ \begin{array}{ll} \left.
\begin{array}{l} -\frac{1}{2}\left( \left( \frac{1}{2}
\gamma_{a,\frac{1}{2}} \frac{1}{\Delta} - \gamma_{i,\frac{1}{2}}
\ln \Delta \right) - \omega_{i,\frac{1}{2}}
\left(\frac{h}{T}\right)^2 \ln \Delta \right) \frac{T^2}{T_i},~~~~~
{}~~~~~~~~~~~~
T \ll T_a=\Delta T_i \\ -\frac{T}{2}\ln 2 + \frac{1}{2}
\left(\gamma_{i,\frac{1}{2}} + \omega_{i,\frac{1}{2}}
\left(\frac{h}{T}\right)^2 \right) \frac{T^2}{T_i} \ln
\frac{T}{T_i} + {\cal O}(\frac{T^2}{T_i}) +  {\cal O}(\Delta
\frac{T^2}{T_i}) ,~~ T_a \ll T \ll T_i \end{array}
\right\} &
\Delta \ll 1 \\ -\frac{1}{2}\left(1 - \frac{\delta}{\Delta}
\right)
\left( \gamma_{a,\frac{1}{2}} + \omega_{a,\frac{1}{2}}
\left(\frac{h}{T}\right)^2  + {\cal O}(\Delta e^{-\Delta}) \right)
\frac{T^2}{T_a},~~~~~~~~~~~~~~~~~~ T \ll \frac{T_a}{\Delta}, & \Delta
 \gg 1
\end{array},
\right.
\end{eqnarray*}
where
$\gamma_{a,\frac{1}{2}} = \pi/6$,
$\omega_{a,\frac{1}{2}} = 1/\pi$ are the
coefficients for the specific heat and susceptibility of the
$S=1/2$ single channel Kondo model, and
$\gamma_{i,\frac{1}{2}} = 1/4$, $\omega_{i,\frac{1}{2}}
= 2/\pi^2$ are those of the $S=1/2$ two-channel
isotropic model, and $\delta$ is a constant close to 1.

The free energy yields the susceptibility and the specific
heat. We find, for any $\Delta>0$, a
linear
specific heat and a temperature independent susceptibility.
Therefore, any small
amount of anisotropy moves the system away from the isotropic
fixed point. In particular,  the $\frac{1}{2}\ln 2$
contribution to the $T=0$ entropy that appears in the isotropic
case \cite{adkondo} is no longer
present. However, for small anisotropy the behavior of
the system in the temperature range $T_a = \Delta T_i < T < T_i$ is
identical to that of the isotropic case, and the isotropic NFL
behavior reemerges upon setting
$\Delta=0$.
For large anisotropy, on the other hand, the leading terms of the
thermodynamic functions are  those of the $S=1/2$ single channel
model with $T_a$ playing the role
of the Kondo temperature.

Computing the ratio $R$,
\begin{eqnarray*}
R_{\frac{1}{2}} = \frac{\chi^i/\chi^e}{C_v^i/C_v^e} =
\frac{\pi^2}{3} \frac{\chi^i}{TC_v} \sim \left\{ \begin{array}{ll}
\frac{8}{3},& \Delta = 0 \\
-8\Delta \ln \Delta, & 0<\Delta \ll 1  \\ 2 + {\cal O}(\Delta e^{-
\Delta}), & \Delta \gg 1 \end{array} \right.
\end{eqnarray*}
we again conclude that turning on the
anisotropy destroys the NFL behavior of the isotropic
model, while the $\Delta=\infty$ single channel FL is only weakly
modified upon reducing the anisotropy.

We now turn to $S=1$. The impurity free energy and $R$ have the
following low-$T$ expressions:
\begin{eqnarray*}
F^i_{S=1} \sim \left\{ \begin{array}{ll} - \frac{1}{2} \left(
\left(
\gamma_{i,1} + \omega_{i,1} \left(\frac{h}{T}\right)^2 \right) -
\left( \alpha + \beta \left(\frac{h}{T}\right)^2  \Delta \ln \Delta
\right) e^{-\frac{1}{\Delta}} \right) \frac{T^2}{T_i}, ~~~ T \ll \Delta
T_i, & \Delta \ll 1 \\ \left. \begin{array}{lc} - \left( \left( 1 +
\frac{\ln
\Delta}{\Delta}
\right) \left( \gamma_{a,\frac{1}{2}} + \omega_{a,\frac{1}{2}}
\left(\frac{h}{T}\right)^2\right) + {\cal O}(e^{-\Delta}) \right)
\frac{T^2}{T_i}, & T \ll T_i, \\ -(\alpha'+\beta'\left( \frac{h}{T}
\right)^2)\frac{1}{\Delta}) \frac{T^2}{T_i} -
\left( \frac{h}{T} \right)^2\frac{c T}{\ln \frac{T}{T_i}}, &
T_i \ll T \ll \Delta T_i \end{array}
\right\} &
\Delta \gg 1
\end{array} \right.
\end{eqnarray*}
\begin{eqnarray*}
R_1 \sim \left\{ \begin{array}{ll} \frac{8}{3} + {\cal O}(e^{-
\frac{1}{\Delta}}), & \Delta \ll 1 \\ 2 + {\cal O}( e^{-
\Delta}), & \Delta \gg 1 \end{array} \right.,
\end{eqnarray*}
where now $\alpha$, $\alpha'$, $\beta$, $\beta'$, and $c$ are
constants of order 1,
$\gamma_{i,1}= \pi/2$ and $\omega_{i,1} = 4/\pi^2$
are the  coefficients for the specific heat and the
susceptibility of the $S=1$ isotropic model.

When $T_i=0$ and $T_a$ is
finite, the system
behaves as a single channel Kondo model with $S=1$, describing at low
temperatures
 a partially screened spin. As
$J_1$ is turned on, ($\Delta \gg 1$),
 the system undergoes another Kondo screening, as
can be seen in the temperature range $ T_i \ll T \ll \Delta T_i=T_a$,
and ends up in the infrared
as a combination of two screened
$S=1/2$ single channel models. The screened $S=1$ isotropic system,
reached as $\Delta=0$, differs in essential ways
  \cite{aj} from the above combination.

We also carried out a numerical solution over the whole
range of anisotropy and temperature (Fig. 1).
Note (e.g. in the spin-1/2 case) the two stage quenching of the
$\ln
2$ high temperature entropy when $T_a \leq T_i$. For
$T_a > T_i$
the quenching occurs in one step at $T_a$. Similarly, the two peaks
present in the
specific
heat for $\Delta \leq 1$ merge and move with $T_a$ as $\Delta$ is
increased past 1. Related observations apply also in the spin-1
case as $\Delta$ is reduced past 1.

We proceed now to interpret our results in terms of fixed-point
hamiltonians that capture the physics of the model in the infrared
limit. Clearly, we reach two different fixed points, the isotropic
$i$, and the anisotropic $a$ when we set
$\Delta=0$, and $\Delta=\infty$, respectively. In the screened
case the $a$-fixed point is unstable to turning
on
$J_1$ ($T_i>0$), and flows to the fixed point describing two
screened spin-1/2 models, to be denoted by $g$.
We conjecture
that for anisotropy $\Delta \geq 1$ the model flows to a line of fixed
points connecting $g$ to $i$, while for
smaller anisotropy it flows to $i$.

In the overscreened case, $i$ is unstable, as remarked earlier and
flows to a new fixed point $g'$. We
conjecture that for anisotropy $\Delta \leq 1$ the model flows to a
line of fixed points connecting $g'$ to $a$, while for larger
anisotropy it flows to $a$. In both cases
 the lines of fixed point hamiltonians $H_{fp}(\Delta)$
 describe systems with
linear specific heat and zero entropy, but not necessarily
FL! \cite{aj}.


The structure of the thermodynamic equations reflects the
flavor symmetry breaking and generalizes to any number of flavors.
The various patterns of the
 $SU(f)$-symmetry breaking  and their relative strengths
will be parametrized by energy scales $T_{n}=D
g_{n}(J_1...J_f),\; \alpha=1,..,f$. The scales set the
excitation
energies and momenta and appear in the thermodynamic equations,
\begin{eqnarray}
\begin{array}{lr} \ln \eta_n = -\frac{T_n}{T} e^{\xi} +
G(\ln(1+\eta_{n-1})+
\ln(1+\eta_{n+1} )), & n \leq f
\nonumber \\
\ln \eta_n = G(\ln(1+\eta_{n-1})+\ln(1+\eta_{n+1})),
&  n>f \nonumber \end{array}
\end{eqnarray}
with $\eta_0\equiv0$.
As an illustration we discuss
the
3-channel problem, with scales $T_1,T_2,T_3$. The
isotropic case is characterized by
$T_3 >0, T_1=T_2=0$. The low energy physics is  NFL for
 $S<3/2$ with $C_v\sim
T^{4/5}$  and  zero-temperature
entropy ${\cal S}=\ln ((\sin (2S+1)\pi /5)/(\sin \pi/5)$.  When the
symmetry is broken to $SU(2) \times U(1)$ a new scale appears.
When $T_1>0,~ T_2=0$ the
models with $S=1/2$ and $S=3/2$ show linear $T$
dependence in
$C_v$, and constant $\chi$, with different coefficients for the
$T_1>T_3$ and $T_1<T_3$ regions. The $S=1$ case is more interesting,
since there
is a change from linear $T$  dependence to NFL behavior, and a
residual
$T=0$ entropy emerges as the parameters
of the system are changed from $T_1<T_3$ to $T_3<T_1$. When $T_1=0,~
T_2>2$
and $S=1/2$, we
have two different NFL fixed points for $T_2<T_3$ and $T_3<T_2$
with
different values of the residual entropy; the $S=1,3/2$
cases have
constant $C_v/T$ and $\chi$. Finally, when all the scales are
finite, the system has constant $C_v$ and $\chi$ for
$S=1/2,1,3/2$, with the coefficients
depending on the relation between the $T_i$ and with intermediate
temperature regions where the properties of the thermodynamic
quantities correspond to those described in the previous cases.
These considerations generalize for any number of flavors $f$.

In forthcoming work we
 study the competition between strength of the
couplings
and the size of symmetry breaking. We also consider the role of the
generalized fusion mechanism in the 2-impurity Kondo model, and
in
models (e.g. with $O(N)-$symmetry)
giving rise to new types of fixed points. Finally, we shall discuss
some phenomenological implications.

We received two preprints \cite{go} addressing the
model by means of Bosonization and the Anderson-Yuval Approach. While
there is a considerable overlap with our work, there are also
interesting differences.

We wish to thank P. Coleman, M. Douglas, A. Gogolin, J.
Moreno, and A. Ruckenstein  for
enlightening  and stimulating discussions. N.A. wishes to thank P.
Woelfle and T. Kopp for their warm hospitality at the Institut fuer
Theorie
der Kondensierten Materie where part of the research was carried
out.

\begin{figure}[here]
\caption{(a) Entropy, (b) Specific heat, and (c) Zero field
susceptibility for
$S=1/2$ as a
function of $T/T_i$, for different values of the anisotropy
parameter
$\Delta=T_a/T_i$.
(d), (e) Same as (a), (b), now for $S=1$. (f) Zero
field susceptibility at $T=0$ as a function of the anisotropy
parameter $\Delta$.}
\label{fig12}
\end{figure}
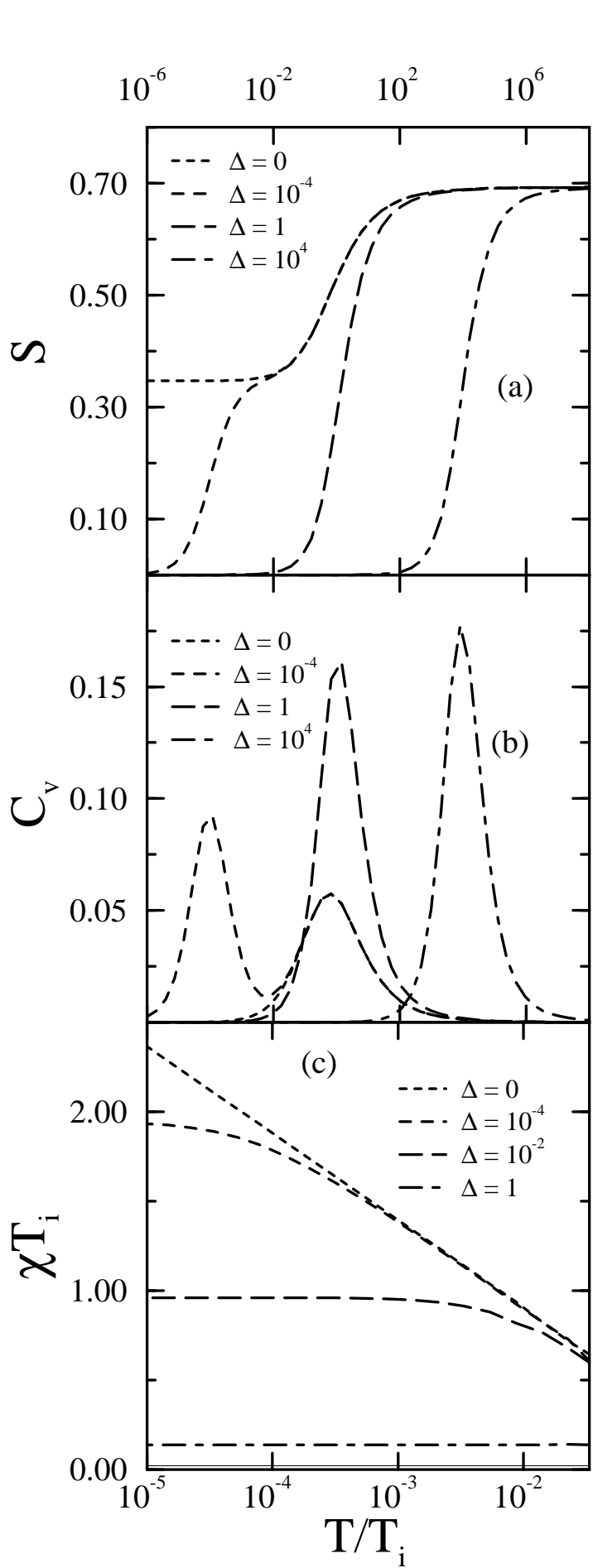
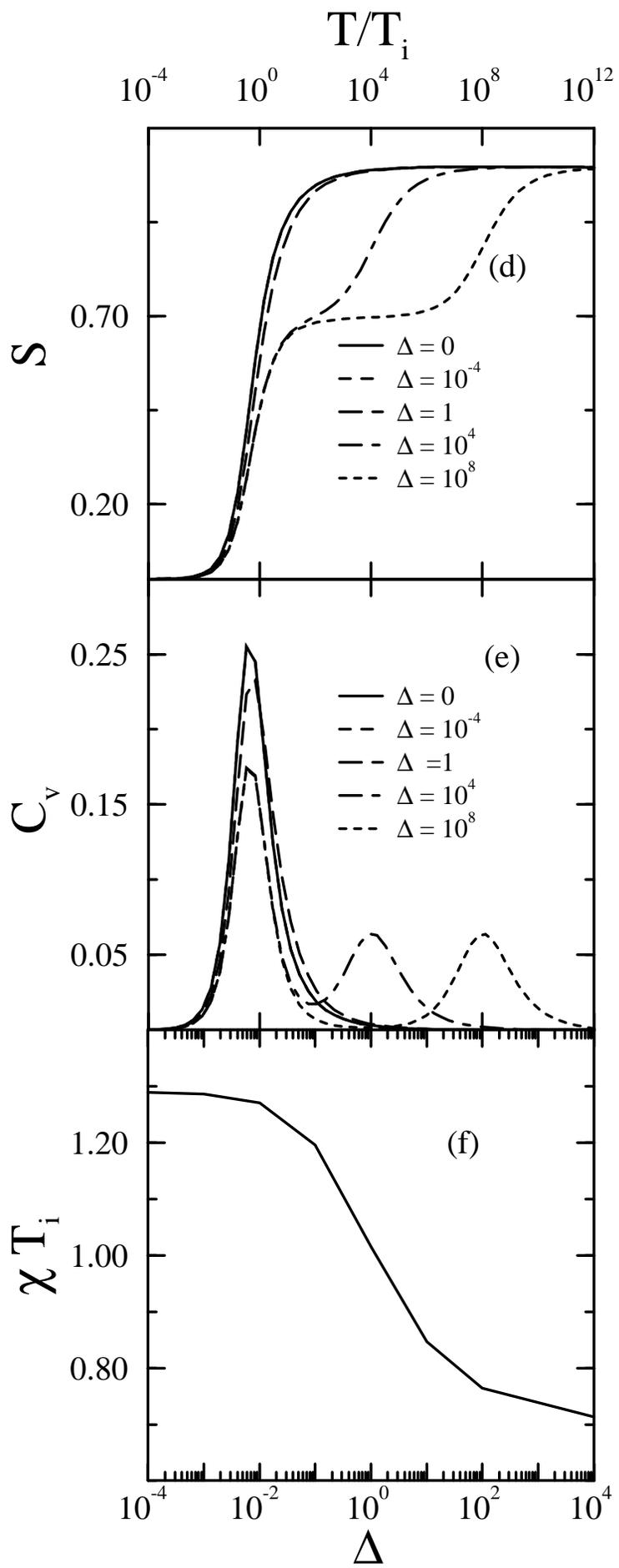
\end{document}